# Synthesis of Colloidal $Mn^{2+}$:ZnO Quantum Dots and High-$T_C$ Ferromagnetic Nanocrystalline Thin Films

Nick S. Norberg,[a] Kevin R. Kittilstved,[a] James E. Amonette,[b] Ravi K. Kukkadapu,[b] Dana A. Schwartz,[a] and Daniel R. Gamelin[a],*

Department of Chemistry, University of Washington, Seattle, WA 98195-1700, and Pacific Northwest National Laboratory, Richland, WA 99352

*To whom correspondence should be addressed: e-mail Gamelin@chem.washington.edu
[a]University of Washington, [b]Pacific Northwest National Laboratory

*Abstract:* We report the synthesis of colloidal $Mn^{2+}$-doped ZnO ($Mn^{2+}$:ZnO) quantum dots and the preparation of room-temperature ferromagnetic nanocrystalline thin films. $Mn^{2+}$:ZnO nanocrystals were prepared by a hydrolysis and condensation reaction in DMSO under atmospheric conditions. Synthesis was monitored by electronic absorption and electron paramagnetic resonance (EPR) spectroscopies. $Zn(OAc)_2$ was found to strongly inhibit oxidation of $Mn^{2+}$ by $O_2$, allowing the synthesis of $Mn^{2+}$:ZnO to be performed aerobically. $Mn^{2+}$ ions were removed from the surfaces of as-prepared nanocrystals using dodecylamine to yield high-quality internally doped $Mn^{2+}$:ZnO colloids of nearly spherical shape and uniform diameter (6.1 ± 0.7 nm). Simulations of the highly resolved X- and Q-band nanocrystal EPR spectra, combined with quantitative analysis of magnetic susceptibilities, confirmed that the manganese is substitutionally incorporated into the ZnO nanocrystals as $Mn^{2+}$ with very homogeneous speciation, differing from bulk $Mn^{2+}$:ZnO only in the magnitude of *D*-strain. Robust ferromagnetism was observed in spin-coated thin films of the nanocrystals, with 300 K saturation moments as large as 1.35 $\mu_B/Mn^{2+}$ and $T_C$ > 350 K. A distinct ferromagnetic resonance signal was observed in the EPR spectra of the ferromagnetic films. The occurrence of ferromagnetism in $Mn^{2+}$:ZnO and its dependence on synthetic variables are discussed in the context of these and previous theoretical and experimental results.

# I. Introduction

Diluted magnetic semiconductors (DMSs)[1] are attracting increasing attention in the physics community due to recent predictions[2,3] and reports[4-6] of room-temperature ferromagnetism in some of these materials. Ferromagnetic DMSs[4-7] have been proposed as pivotal components in a new category of spin-based electronics (spintronics)[8,9] devices that aim to control electron spin currents as well as charge currents to increase data processing speeds, reduce power consumption, reduce hardware dimensions, and possibly introduce new functionalities to semiconductor information processing technologies.

Theoreticians have identified ZnO as an excellent candidate host semiconductor for supporting high-Curie-temperature (high-$T_C$) ferromagnetism when doped with a variety of 3d transition metal ions, particularly $Mn^{2+}$.[2,3] Experimentalists have verified these predictions in some cases, with ferromagnetism above room temperature (RT) reported for thin films of ZnO doped with $Co^{2+}$,[10] $Co^{2+}/Fe^{2+}$,[11] and $V^{2+}$,[12] prepared by vacuum deposition methods such as pulsed laser deposition (PLD). These findings remain controversial, however, and several laboratories have claimed to observe ferromagnetism arising only from phase segregated impurities and not from the DMSs themselves.[13,14] Not until very recently was ferromagnetism above RT reported for $Mn^{2+}$:ZnO,[15] even though this DMS was specifically highlighted in theoretical studies for its high-$T_C$ ferromagnetism potential.[2,3] Many earlier studies of $Mn^{2+}$:ZnO revealed only paramagnetism, or at best ferromagnetism below RT, with $T_C$ ranging from 37 K to 250 K, and many have low saturation moments indicative of only partial magnetic ordering.[10,16-18] The wide range of magnetic properties displayed by this and other ZnO DMSs prepared by apparently similar methods[10,19] suggests that the criteria necessary for ferromagnetism are highly sensitive to the preparation conditions and ultimately to the materials composition. Developing reproducible methods for preparing high-$T_C$ ferromagnetic DMSs is essential for their use in spintronics technologies but remains a central challenge in this field. By addressing this challenge, the understanding of the fundamental origins of this interesting magnetic behavior will also be advanced.

Although ferromagnetism in DMSs has attracted attention primarily from physicists, we envision chemistry as offering several advantages in this research area. We speculate that direct chemical syntheses of ZnO DMSs can provide better control over materials composition than is obtained with some high-temperature vacuum deposition and solid-state synthetic techniques, which often use dopant source materials that are themselves undesirable contaminants (e.g. manganese oxides for $Mn^{2+}$:ZnO,[15,17] NiO for $Ni^{2+}$:ZnO,[20] and cobalt oxides or Co metal for $Co^{2+}$:ZnO[21]) or require high temperatures and reducing conditions that may promote segregation of metallic precipitates.[13,18] The advantage of the direct chemical approach is illustrated in the case of $Ni^{2+}$:ZnO. The solid solubility of $Ni^{2+}$ in ZnO is very low[22] and there is a large driving force for phase segregation. $Ni^{2+}$:ZnO prepared by PLD using NiO as the source of $Ni^{2+}$ showed a magnetic hysteresis only below 5 K[20] that resembled the characteristic magnetic behavior of nanoscale NiO,[23] raising concerns over the possibility of nanoscale NiO contaminants. In contrast, $Ni^{2+}$:ZnO DMSs prepared from ionic solutions showed ferromagnetism above room temperature ($T_C > 350K$) that was demonstrably an intrinsic property of the $Ni^{2+}$:ZnO DMS.[24]

In this paper we report the direct chemical synthesis of high-quality colloidal $Mn^{2+}$:ZnO quantum dots (QDs) and the use of these nanocrystals as solution-phase precursors for the preparation of nanocrystalline thin films by spin-coat processing. Many proposed spintronics devices involve DMSs of nanometer dimensions,[8] and solution syntheses provide the opportunity to study freestanding DMS crystals in this size regime. These syntheses can be scaled up easily to produce gram quantities of the desired DMS, which greatly facilitates rapid exploration of the physical properties of these materials. Colloidal nanocrystals also have accessible surface chemistry that allows functionalization and processing in a variety of solvents, and this opens the door to potential building-block applications in nanotechnology, an area in which self-assembly is of increasing importance. Colloidal $Mn^{2+}$-doped DMS nanocrystals of ZnS,[25] CdS,[26] ZnSe,[27] CdSe,[28] PbSe,[29] and InAs[30] have been reported previously. The synthesis and characterization of colloidal $Co^{2+}$- and $Ni^{2+}$-doped ZnO DMS-QDs,[24,31,32] and nanocrystalline $Mn^{2+}$:ZnO powders prepared from solution[33] have also been described recently. While this paper was under review, a report of $Mn^{2+}$-doped ZnO colloids appeared.[34] X-ray absorption and EPR spectroscopies were used to demonstrate that $\geq$ ~45% of the manganese was segregated at the nanocrystal surfaces in this preparation.

The synthesis of $Mn^{2+}$:ZnO nanocrystals in this study was monitored using electron paramagnetic resonance (EPR) and electronic absorption spectroscopies. A method for removing $Mn^{2+}$ ions from the nanocrystal surfaces is presented, and its efficacy is demonstrated by EPR spectroscopy. Simulations of the experimental X- and Q-band EPR spectra confirm substitutional doping of the $Mn^{2+}$ ions in ZnO with very homogeneous speciation and with bulk-like $Mn^{2+}$ ground-state electronic structures. These colloidal DMS nanocrystals represent highly versatile solution-phase building blocks that are compatible with numerous standard solution processing methods. We report robust, high-$T_C$ ferromagnetism in thin films of these nanocrystals prepared by spin coating, with 300 K saturation moments as high as 1.35 $\mu_B/Mn^{2+}$, nearly an order of magnitude higher than that previously reported.[15] The $Mn^{2+}$:ZnO ferromagnetism is accompanied by a broad ferromagnetic resonance EPR signal that reflects the high multiplicity of the ferromagnetic ground state. The preparation of more complex ferromagnetic semiconductor nanoarchitectures by solution processing methods may therefore be envisioned.

## II. Experimental

**A. Materials.** Zinc acetate dihydrate ($Zn(OAc)_2 \cdot 2H_2O$ (98%, <0.0005% magnetic impurities)) (Strem), manganese acetate tetrahydrate ($Mn(OAc)_2 \cdot 4H_2O$, 99.999%) (Strem), manganese nitrate hydrate ($Mn(NO_3)_2 \cdot xH_2O$, 98%) (Sigma-Aldrich), tetramethylammonium hydroxide ($N(CH_4)_4OH \cdot 5H_2O$, 97%) (Sigma-Aldrich), dodecylamine ($CH_3(CH_2)_{11}NH_2$, 98%) (Sigma-Aldrich), trioctylphosphine oxide (TOPO, $((CH_3(CH_2)_7)_3P(O)$, 90%)) (Sigma-Aldrich), dimethylsulfoxide (DMSO, 99.7%) (Acros), and absolute ethanol (AAPER) were purchased and used as received.

**B. Sample Preparation.** $Mn^{2+}$:ZnO nanocrystals were synthesized at room temperature by dropwise addition of 1.7 equivalents of an ethanolic solution of 0.55 M tetramethylammonium hydroxide ($N(Me)_4OH \cdot 5H_2O$) to a 0.10 M solution of $xMn(OAc)_2 \cdot 4H_2O/(1-x)Zn(OAc)_2 \cdot 2H_2O$ dissolved in DMSO under constant stirring. Following initial rapid growth from solution that proceeds on the timescale of minutes,[32] growth of the nanocrystals continues by Ostwald ripening on the timescale of several days at room temperature or can be accelerated by heating to ca. 60 ˚C. After growth, the $Mn^{2+}$:ZnO nanocrystals were precipitated by addition of ethyl acetate and resuspended in ethanol. Iterative washing by precipitation with heptane and resuspension in ethanol ensures that excess reactants have been removed from the product. After washing, the nanocrystals suspended in ethanol were capped with dodecylamine and resuspended in toluene. A sample in which $Mn^{2+}$ ions were deliberately bound to the surfaces of ZnO nanocrystals was prepared by first synthesizing pure ZnO nanocrystals following the above procedure with no added $Mn(OAc)_2$. After washing and resuspending the nanocrystals in ethanol, a small amount of $Mn(OAc)_2$ (ca. 2% of $Zn^{2+}$) was mixed into the ZnO suspension and 0.002 equiv of an ethanolic solution of LiOH were added dropwise. The resulting surface-bound $Mn^{2+}$:ZnO nanocrystals were subsequently washed, capped with dodecylamine, and resuspended in toluene for EPR experiments.

To remove $Mn^{2+}$ ions from the nanocrystal surfaces, dodecylamine-capped nanocrystals were heated in dodecylamine (m.p. ca. 30 ˚C) at 180 ˚C for ca. 30 minutes under nitrogen. The nanocrystals were allowed to cool to below 80 ˚C, and were then precipitated and washed with ethanol. The resulting amine-treated powders were resuspended in toluene or other nonpolar solvents to form colloidal suspensions of high optical quality that are stable for several months. In some experiments, $Mn^{2+}$ ions were removed from the nanocrystal surfaces by heating in technical grade trioctylphosphine oxide (TOPO) following a procedure described previously.[32]

Nanocrystalline thin films of $Mn^{2+}$:ZnO were prepared by spin-coating amine-treated $Mn^{2+}$:ZnO colloids onto 1 cm x 0.5 cm fused-silica substrates. The films were annealed at 525 ˚C for 2 minutes in air after each spin-coated layer was added. Film A comprised 40 coats and had a mass of 0.59 mg. Films B and C were made from 20 coats each and had masses of 0.24 and 0.28 mg, respectively.

**C. Physical Characterization.** X-band (9.34 GHz) electron paramagnetic resonance (EPR) spectra were collected on a Bruker EMX EPR spectrometer at the University of Washington and Q-band (34.0 GHz) EPR spectra were collected on a Bruker ESP300E EPR spectrometer at the Pacific Northwest National Laboratories. The X- and Q-band EPR data for the freestanding nanocrystals were recorded at room temperature on colloids suspended in toluene. Room-temperature X-band EPR spectra of thin films were measured by attaching the films to EPR tubes with non-magnetic tape and aligning them perpendicular to the magnetic field in the EPR cavity. Simulations of the X- and Q-band EPR spectra were performed using a full matrix diagonalization routine implemented by SIM (ver.2002), written and graciously provided by Prof. Høgni Weihe of the University of Copenhagen.

Absorption spectra of freestanding nanocrystals were measured at room temperature on colloids using 1 cm pathlength cuvettes using a Cary 500E (Varian) spectrophotometer. To observe both weak and intense absorption features during a base titration experiment, a portion of the sample was extracted from the reaction vessel, measured, diluted by a constant factor (70x), and then re-measured. Low-temperature MCD spectra were collected on drop-coated films (frozen solutions) as described previously.[32] MCD intensities were measured as the differential absorbance $\Delta A = A_L - A_R$ (where L and R refer to left and right circularly polarized photons) and are reported as $\theta(deg) = 32.9 \Delta A/A$. Colloid luminescence spectra were measured in 1 cm x 1 cm fluorescence cuvettes on a Jobin Yvon FluoroMax-2 Fluorimeter.

Cu $K_\alpha$ X-ray powder diffraction data were collected on a Philips PW 1830 X-ray diffractometer for powders and on a Rigaku Rotaflex RTP300 diffractometer for thin films. High resolution transmission electron microscopy (HRTEM) images were collected at the Pacific Northwest National Laboratories on a JOEL 2010 transmission electron microscope (200 kV) with a high brightness $LaB_6$ filament as an electron source. Dopant concentrations were determined by inductively coupled plasma atomic emission spectrometry (ICP-AES, Jarrel Ash model 955).

Magnetic susceptibility data for $Mn^{2+}$:ZnO nanocrystalline powders rapidly precipitated from toluene and for spin-coated thin films were collected using a Quantum Design MPMS-5S SQUID magnetometer. All data were corrected for the diamagnetism of the substrate and sample holder.

**III. Results**

Figure 1a shows electronic absorption spectra of a 0.10 M DMSO solution of 2% Mn(OAc)$_2$·4H$_2$O/98% Zn(OAc)$_2$·2H$_2$O collected following successive additions of 0.15 equiv of 0.55 M N(Me)$_4$OH in ethanol. An intense absorption feature appears at ca. 30000 cm$^{-1}$ after addition of 0.45 equiv of base that is readily identified as the first excitonic band gap transition of ZnO.[35,36] Further base addition yields an approximately linear increase in band gap absorbance. A sub-bandgap feature at 24000 cm$^{-1}$ also appears after 0.45 equiv of base have been added and its intensity grows with continued addition of base, giving the colloidal suspension a brown color. This sub-bandgap feature has been observed previously in other Mn$^{2+}$:ZnO preparations.[17,22,37,38] The solid line in Figure 1a shows the spectrum collected after 1.65 equiv of base were added. Base addition beyond ca. 1.65 equiv caused the suspensions to begin to cloud.

Figure 1b shows absorption spectra of 0.20% Mn$^{2+}$:ZnO colloids after heating in dodecylamine to remove Mn$^{2+}$ ions from the nanocrystal surfaces (vide infra). These nanocrystals, prepared with an initial dopant level of 0.50% Mn$^{2+}$, were found by ICP-AES to contain 0.20 ±0.01% Mn$^{2+}$ (i.e. Zn$_{0.998}$Mn$_{0.002}$O). Figure 1b also shows the absorption spectrum of a thin film prepared by spin coating the same colloids onto a fused silica substrate. We refer to this film throughout this paper as Film A. The band gap energy of the nanocrystalline thin film is similar to that of bulk ZnO (ca. 27000 cm$^{-1}$ at 300 K).[6] The band gap energy of the colloids determined by the same method is slightly larger, attributable to quantum confinement. The sub-bandgap absorption feature is observed in both the colloids and the thin film.

Figure 2 shows room-temperature X-band EPR spectra of colloidal Mn$^{2+}$:ZnO nanocrystals in toluene. Figure 2a is the spectrum of ZnO nanocrystals with Mn$^{2+}$ ions deliberately bound to the surfaces. Figures 2b-d show EPR spectra of aliquots removed from a reaction in which 1.7 equiv of 0.55 M N(Me)$_4$OH were added to a 0.10 M DMSO solution of 0.02% Mn(OAc)$_2$/99.98% Zn(OAc)$_2$. Samples for EPR measurements were removed from the reaction (b) 10 minutes after base addition at room temperature, (c) after allowing nanocrystal growth for 2 hours at 60 °C, and (d) after washing the nanocrystalline product and heating in dodecylamine for 30 minutes at 180 °C. All samples were washed and capped with dodecylamine. The average nanocrystal diameters for (b) and (c) were estimated from the nanocrystal band gap absorption energies[36] to be 4.0 and 5.6 nm, respectively. All spectra show six main features, characteristic of the hyperfine coupling of $Mn^{2+}$ (I = 5/2). The spectra evolve from having six broad signals with an apparent hyperfine splitting of ca. 88 Gauss (Figure 2a) to showing extensive resolved fine structure with narrow features and an apparent hyperfine splitting of ca. 78 Gauss for the major features (Figure 2d).

Figure 3a presents powder X-ray diffraction (XRD) data for the 0.20% $Mn^{2+}$:ZnO nanocrystals shown in Figure 1b precipitated rapidly from toluene (dashed), and for Film A (solid), prepared by spin-coating the same nanocrystals onto fused silica. All of the peaks in the two diffraction patterns match those of wurtzite ZnO, shown as the indexed lines in the bottom of Figure 3a. TEM data collected for the same nanocrystals are shown in Figure 3b-d. The images in Figures 3b and 3d show approximately spherical and highly crystalline nanocrystals. The characteristic lattice spacings of wurtzite ZnO are readily identified in the high resolution TEM images of these nanocrystals (Figure 3d). Analysis of ca. 100 nanocrystals (Figure 3c) yields an average crystal diameter of 6.1 ± 0.7 nm.

Figure 4 shows absorption data collected for various solutions of 0.002 M $Mn(OAc)_2 \cdot 4H_2O$ in DMSO that were allowed to stand open to the atmosphere for several days. This $Mn^{2+}$ concentration is the same as that used for the synthesis of 2% $Mn^{2+}$:ZnO nanocrystals. The inset shows the change in absorption of the 0.002 M $Mn(OAc)_2 \cdot 4H_2O$ solution when exposed to air for 5 days. The solution turns brown and a broad tailing absorption feature emerges. The open circles in Figure 4 show the absorption intensity at 20000 $cm^{-1}$ after 5 days for solutions of 0.002 M $Mn(OAc)_2 \cdot 4H_2O$ with varying amounts of $Zn(OAc)_2 \cdot 2H_2O$ added. A sharp decrease in the formation of the brown absorption intensity is observed as the ratio of $Zn^{2+}$ to $Mn^{2+}$ increases, reaching almost zero at a ratio of 3 $Zn(OAc)_2 \cdot 2H_2O$ to 1 $Mn(OAc)_2 \cdot 4H_2O$. The absorption intensities measured for analogous 0.002 M $Mn(OAc)_2 \cdot 4H_2O$ solutions allowed to stand anaerobically (■) or in air with 0.010 M NaOAc added (▲) are also included in Figure 4. Addition of NaOAc causes 50% more absorption intensity after 5 days, whereas the anaerobic solution shows only a small increase in absorption after 5 days. When $Mn(NO_3)_2 \cdot xH_2O$ (♦) is used in place of $Mn(OAc)_2 \cdot 4H_2O$, little or no absorbance change is observed after 5 days in air.

Room-temperature X- and Q-band EPR spectra of surface-cleaned $Mn^{2+}$:ZnO colloidal nanocrystals prepared from a 0.02% $Mn(OAc)_2 \cdot 4H_2O$/99.98% $Zn(OAc)_2 \cdot 2H_2O$ solution as in Figure 2d are displayed in Figure 5. Both experimental spectra are highly resolved, exhibiting small peak widths for all features and clearly showing six major hyperfine lines. Many additional features are also present, however, including three outer peaks on either side of the main sextet in the X-band EPR spectrum.

Figure 6 shows 300 K absorption and 5 K MCD spectra of freestanding TOPO-capped 1.1% $Mn^{2+}$:ZnO nanocrystals. The 5 K MCD spectrum shows a broad negative pseudo-$\mathscr{A}$ term MCD feature associated with this sub-bandgap absorption that tails throughout the visible energy window (shown down to 16000 $cm^{-1}$). The MCD intensity becomes negative again at 28000 $cm^{-1}$ with the onset of the ZnO band edge. The pseudo-$\mathscr{A}$ term MCD intensity increases with increasing applied field and approaches saturation above ca. 4 Tesla (Figure 6 inset).

Figure 7 shows room-temperature absorption and luminescence spectra of ZnO, 0.13% $Mn^{2+}$:ZnO (estimated concentration), and 1.3% $Mn^{2+}$:ZnO colloidal nanocrystals capped with dodecylamine and suspended in toluene. All three samples were heated with dodecylamine following the procedure described above for removal of surface-bound $Mn^{2+}$. All luminescence spectra were collected with excitation into the band gap region (28200 $cm^{-1}$). The absorption intensities at the excitation energies, which ranged between 0.1 and 0.4 absorbance units, were normalized in Figure 7, and the emission intensities were changed proportionally. The pure ZnO nanocrystals show a broad visible luminescence band centered at ca. 18600 $cm^{-1}$ and a relatively intense UV emission band at 26900 $cm^{-1}$. The 0.13% $Mn^{2+}$:ZnO colloids show a similar luminescence spectrum but the visible and UV emission intensities have been reduced by 42 and 69%, respectively, relative to the pure ZnO nanocrystal spectrum. The visible emission in the 1.3% $Mn^{2+}$:ZnO colloids is quenched by 96%, and they also do not show the same excitonic emission feature in the UV but show only a weak intensity that may arise from scattering.

Figure 8 shows the magnetic susceptibilities of freestanding 0.20 (±0.01)% $Mn^{2+}$:ZnO nanocrystals and three thin films (Films A, B, and C) prepared by spin coating the 0.20% $Mn^{2+}$:ZnO nanocrystals onto fused silica substrates. Film A is the same film shown in Figure 1b. The magnetization of the freestanding nanocrystals is linear with small applied fields (< 1 Tesla, or 10000 Oe) at 5K and is strongly temperature dependent. In sharp contrast, the thin films all show rapid magnetic saturation and clear magnetic hystereses at both low and high temperatures. For each film, a temperature- and field-dependent magnetization signal similar to that of the freestanding nanocrystals is observed superimposed on the nearly temperature-independent magnetic hysteresis. The 300 K saturation moments ($M_S$) for the three films are (A) 0.67 $\mu_B/Mn^{2+}$, (B) 1.18 $\mu_B/Mn^{2+}$, and (C) 1.35 $\mu_B/Mn^{2+}$. The 300 K remanent magnetization ($M_R$) and coercivity ($H_C$) values are very similar for all three films, with averages of 0.12 ±0.01 $M_S$, and 92 ±7 Oe, respectively. The magnetic hysteresis parameters for Films A, B, and C are plotted as a function of temperature in Figure S2 of Supporting Information. All three hysteresis parameters decrease gradually as the temperature is increased from 5 K to the instrument limit of 350 K.

Zero-field-cooled (ZFC) magnetization measurements for Films A and B from Figure 8 are shown in Figure 9a. Both films show a steep decrease in magnetization with increasing temperature between 5 and 10 K, followed by a gradual increase in magnetization between 10 and 350 K. Figure 9b shows the 5 K linear magnetization signal for Film A obtained by subtraction of the ferromagnetic signal from the 5 K data in Figure 8. These data are compared quantitatively to the data collected for the freestanding nanocrystals, shown in Figure 8. The linear magnetization signal of the thin film is 64% as large as that of the freestanding nanocrystals. Figure 9c plots the temperature dependence of the 1 Tesla magnetization for Film A obtained by subtracting the ferromagnetic signals from the data in Figure 8 and from analogous data collected at intermediate temperatures. These data show a strong decrease in magnetization with increasing temperature, reaching nearly zero at room temperature.

Figure 10 compares the 300 K X-band EPR spectra of Film A and the freestanding 0.20% $Mn^{2+}$:ZnO colloids used to make this film. Whereas the colloid spectrum shows extensive hyperfine structure between 3000 and 3600 Gauss, the thin film EPR spectrum shows an intense, broad resonance spanning the entire spectral range. Hyperfine structure similar to that of the colloid spectrum can be observed superimposed on the broad feature in the spectrum of the thin film. Additionally, a new, sharp resonance at $H = 3362$ Gauss ($g = 2.00$) is observed in the thin film.

## IV. Analysis and Discussion

**A. Manganese incorporation during growth of ZnO nanocrystals.** The synthesis of $Mn^{2+}$-doped ZnO nanocrystals was monitored by electronic absorption spectroscopy. The absorption spectra in Figure 1 show the characteristic ZnO band gap absorbance at ca. 30000 $cm^{-1}$ upon addition of base to the DMSO solution of $Zn^{2+}$, but only after 0.45 equiv of base has been added. This induction period reflects the build-up of $Zn^{2+}$ precursor concentrations toward supersaturation prior to ZnO nucleation, as described by the LaMer model.[39] The precursors for ZnO nucleation under these conditions are believed to be basic zinc acetate clusters.[35,40] After 0.45 equiv, subsequent base addition causes further nucleation of ZnO and a stoichiometric increase in band gap absorption intensity is observed (Figure 1). Similar results were obtained for the synthesis of $Co^{2+}$- and $Ni^{2+}$-doped ZnO nanocrystals.[32]

In addition to the ZnO excitonic absorption, a sub-bandgap absorption feature at 24000 $cm^{-1}$ is observed that also appears only after 0.45 equiv of base has been added, i.e. its appearance coincides with the appearance of ZnO. This sub-bandgap absorption feature is not observed in pure ZnO QDs synthesized under these conditions and it is therefore attributed to the $Mn^{2+}$. The fact that its appearance accompanies ZnO formation provides a strong indication that the $Mn^{2+}$ ions giving rise to this absorption intensity are associated with crystalline ZnO. This intensity is assigned as a charge-transfer transition (see Section IV.F for analysis). Because the $Mn^{2+}$ d-d transitions are extremely weak, it was not possible to extract as detailed information about the nucleation process from these titration experiments as was possible for $Co^{2+}$:ZnO, where the $Co^{2+}$ ligand-field absorption could be used to follow the reaction.[32] Instead, EPR spectroscopy was used to probe the $Mn^{2+}$ ions during synthesis.

The EPR spectra shown in Figure 2b-d were collected at different stages of nanocrystal synthesis under reaction conditions similar to those of Figure 1. They reveal the changing environment of the $Mn^{2+}$ ions during the preparation. For reference, we have prepared pure ZnO quantum dots and then deliberately bound $Mn^{2+}$ ions to their surfaces (i.e. 100% surface $Mn^{2+}$). The resulting EPR spectrum is shown in Figure 2a. The breadth of the features in Figure 2a is attributed to the inhomogeneous $Mn^{2+}$ speciation on the nanocrystal surfaces. The EPR spectrum of $Mn^{2+}$:ZnO nanocrystals collected shortly after addition of base (Figure 2b) generally resembles that of surface-bound $Mn^{2+}$:ZnO, but shows improved resolution of the fine structure. The emergence of structure in this spectrum is attributed to partial incorporation of $Mn^{2+}$ into the ZnO lattice. From the energy of the first excitonic absorption feature, the average nanocrystal diameter for this sample is estimated to be ca. 4.0 nm. At this size, even a statistical distribution of the dopants in the ZnO nanocrystals will result in a high proportion (ca. 25%) of the $Mn^{2+}$ ions at the surfaces of the nanocrystals because of the large surface-to-volume ratios. The spectrum in Figure 2c was collected after allowing the nanocrystals to Ostwald ripen at 60 °C for 2 hours. An average nanocrystal diameter of ca. 5.6 nm is estimated from the red-shifted excitonic absorption spectrum of these colloids (ca. 20% surface). The EPR spectrum of Figure 2c is somewhat sharper and shows slightly better resolved hyperfine structure than that of Figure 2b, reflecting increased homogeneity in the $Mn^{2+}$ speciation. The increased homogeneity thus qualitatively follows the decrease in the surface-to-volume ratio upon increasing the crystal diameters from 4.0 to 5.6 nm. The level of resolved hyperfine structure observed in Figure 2c is similar to that reported previously for calcined $Mn^{2+}$:ZnO nanocrystals.[33]

**B. Removing dopants from nanocrystal surfaces with dodecylamine.** The presence of surface $Mn^{2+}$ ions lowers the quality of the DMS-QDs and may compromise or obfuscate their physical properties. We have previously demonstrated that $Co^{2+}$ and $Ni^{2+}$ ions on ZnO nanocrystal surfaces can be removed by heating the nanocrystals in technical grade TOPO.[32] Here we report an improved procedure for removing surface-bound dopants that involves heating the quantum dots in a solution of dodecylamine at 180 °C for ca. 30 minutes. The dodecylamine ligates surface-exposed dopants and solvates them, leaving only internal dopants within the cores of the colloidal ZnO nanocrystals. The 300 K EPR spectrum of colloidal $Mn^{2+}$:ZnO nanocrystals cleaned in this way (Figure 2d) shows exceptional resolution and rich fine structure. From the excellent agreement between this spectrum and that of bulk $Mn^{2+}$:ZnO (see Section IV.C), we conclude that the resulting nanocrystals contain solely internal $Mn^{2+}$ ions with homogeneous speciation. These data emphasize the importance of removing surface-exposed dopants from DMS-QDs in order to ensure high quality materials suitable for further study and application.

**C. EPR of surface-cleaned $Mn^{2+}$:ZnO nanocrystals.** The room-temperature X- and Q-band EPR spectra of the colloidal 0.02% $Mn^{2+}$:ZnO nanocrystals (Figure 5) were simulated using the axial spin Hamiltonian given in Equation 1.

$$H = g\mu_B \mathbf{H} \cdot \mathbf{S} + A\mathbf{S} \cdot \mathbf{I} + D\left[S_z^2 - \frac{1}{3}S(S+1)\right] \quad (1)$$

The first term describes the Zeeman interaction, the second term describes electron-nuclear magnetic hyperfine coupling ($^{55}$Mn nucleus, $I = 5/2$), and the third term accounts for the axial zero-field splitting caused by the hexagonal symmetry of wurtzite ZnO. The $g$ and $A$ tensors were approximated as isotropic in the simulations since the spectral anisotropy is unresolved in the powder-averaged EPR spectrum.[41] Excellent quantitative agreement with the experimental X- and Q-band spectra could be achieved in simulations (simulations X1 and Q1) using a single set of parameters ($g = 1.999$, $A = -74.0 \times 10^{-4}$ cm$^{-1}$, and $D = -2.36 \times 10^{-2}$ cm$^{-1}$ ($\pm 0.05 \times 10^{-2}$ cm$^{-1}$)). These parameters agree well with literature values ($g_\| = 1.998$, $g_\perp = 2.000$, $A_\| = -74.0 \times 10^{-4}$ cm$^{-1}$, $A_\perp = -73.5 \times 10^{-4}$ cm$^{-1}$, and $D = -2.36 \times 10^{-2}$ cm$^{-1}$) obtained from the study of single crystals of $Mn^{2+}$:ZnO.[42,43] These simulations confirm that the main hyperfine lines in the spectra are due to $\Delta M_S = \pm 1$ transitions having $\Delta M_I = 0$ and that the many additional lines arise from the formally forbidden resonances having $\Delta M_I \neq 0$, which gain intensity from off-diagonal terms that are probed in the powder-averaged spectrum.[42] A small amount of $D$-strain ($\sigma = 2\%$ of $D$) was required for optimal simulation of the experimental data (simulations X1 and Q1). For comparison, simulations X2 and Q2 in Figure 5 do not include $D$-strain. Since $D$ originates in the trigonal ligand-field perturbation of the $Mn^{2+}$ in hexagonal (wurtzite) ZnO, the presence of *D*-strain reflects a greater range of trigonal distortions for the ensemble of $Mn^{2+}$ ions in the nanocrystals than in bulk $Mn^{2+}$:ZnO. This *D*-strain is thus attributable to lattice relaxation effects in the nanocrystals that are not present in bulk $Mn^{2+}$:ZnO. In nanocrystalline $Mn^{2+}$:ZnO, $Mn^{2+}$ ions close to the surface of a nanocrystal will induce greater lattice distortion than those deep in the core because of the crystal's greater capacity to relax structurally near its surface. This *D*-strain is small (2%), however, and the close match between the simulated and experimental spectra confirms that the $Mn^{2+}$ ions are substitutionally doped into the ZnO nanocrystals with very homogeneous speciation. We note that, except for the slight *D*-strain, the ground-state electronic structure of $Mn^{2+}$ in ZnO nanocrystals is indistinguishable from that in bulk ZnO to a high degree of accuracy. This latter point does not support earlier claims that dopant-host electronic interactions may be dramatically enhanced in $Mn^{2+}$ DMS nanocrystals and lead to large changes in the electronic structures of the $Mn^{2+}$ ions,[44] although the DMS nanocrystals studied here are too large to be strongly quantum confined (Section IV.D).

**D. Structural Characterization of the $Mn^{2+}$:ZnO Nanocrystals.** In addition to solvating surface dopant ions, heating the nanocrystals in dodecylamine also induces further growth by Ostwald ripening. The absorption spectrum of amine-treated 0.20% $Mn^{2+}$:ZnO nanocrystals in Figure 1b shows a bandgap that is clearly smaller than that of the initially prepared QDs in DMSO shown in Figure 1a. The slight quantum confinement apparent from Figure 1b suggests that the nanocrystal diameters are between 6 and 7 nm (ZnO Bohr radius = 3.5 nm), although estimates of ZnO nanocrystal sizes by band gap measurements are not particularly reliable in this size range. Analysis of the broadened XRD peaks observed for these 0.20% $Mn^{2+}$:ZnO nanocrystals (Figure 3a) using the Scherrer equation yields an average nanocrystal diameter of 6.1 nm, consistent with the absorption data. Similarly, TEM images of the same nanocrystals (Figure 3b-d) confirm the crystallinity and pseudo-spherical shapes of these crystals, and yield an average crystal diameter of 6.1 ± 0.7 nm.

**E. $Zn(OAc)_2$ inhibition of $Mn(OAc)_2$ oxidation in DMSO.** $Mn^{2+}$ is easily oxidized in air to form oxides such as $Mn_2O_3$ and $Mn_3O_4$, and identification of appropriate reducing conditions that allow the preparation of $Mn^{2+}$-doped ZnO by high-temperature methods has been extensively investigated.[45] To maintain MnO as divalent $Mn^{2+}$ at 900 °C, for example, $O_2$ pressures below $10^{-7}$ atm are required. Similarly, $Mn^{2+}$ solutions are sensitive to oxidation in air at room temperature, and stock solutions of $Mn(OAc)_2$ in DMSO stored under air will turn from colorless to brown over a period of several days. Absorption spectra of a solution of 0.002M $Mn(OAc)_2 \cdot 4H_2O$ in DMSO collected immediately after preparation and again after 5 days are shown in the inset of Figure 4. The brown color arises from a broad, tailing absorption feature having a shoulder at 20000 $cm^{-1}$, clearly distinguishable from the substitutional $Mn^{2+}$:ZnO sub-bandgap absorption at 24000 $cm^{-1}$ (Figure 1). A control experiment in which a similar $Mn^{2+}$ solution was prepared and stored anaerobically showed 84% less discoloration after 5 days (Figure 4, ■), verifying that the discoloration involves oxidation of $Mn^{2+}$ by atmospheric $O_2$. Basic solutions of $Mn^{2+}$ are known to convert readily to a variety of complex oxides and hydroxides under air,[46] such as described by Equation 2, and many of these oxidation products are deeply colored.

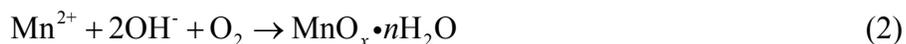

$$Mn^{2+} + 2OH^- + O_2 \rightarrow MnO_x \bullet nH_2O \qquad (2)$$

During our studies of the synthesis of $Mn^{2+}$:ZnO from solution, we observed that similar stock solutions of $Mn^{2+}$ that also contained $Zn(OAc)_2$ do not change color in air after several days. This observation suggests that $Zn(OAc)_2$ inhibits the oxidation of $Mn^{2+}$. Data collected for a range of $Zn(OAc)_2 \cdot 2H_2O$ concentrations (Figure 4) reveal that inhibition of $Mn^{2+}$ oxidation is strongly dependent on the concentration of $Zn(OAc)_2$, with almost complete inhibition achieved at Zn:Mn stoichiometries as low as only ca. 3:1 (compared to the $Mn^{2+}$:ZnO synthesis conditions of >49:1 used here). In other control experiments, 0.002M $Mn(NO_3)_2$ was stored under air for 5 days and showed almost no color change (Figure 4, ♦), but addition of Na(OAc) to a 0.002M $Mn(OAc)_2$ solution resulted in a substantial increase in discoloration (Figure 4, ▲), suggesting that the acetate actually promotes the oxidation of $Mn^{2+}$. These data are consistent with the conclusion that the inhibition of $Mn^{2+}$ oxidation by $Zn(OAc)_2$ is due to pH effects. Although acetate is a weak base, it is apparently able to accelerate the oxidation of $Mn^{2+}$, which should be strongly pH dependent as described by Equation 2. Nitrate, an even weaker base, is not as capable of accelerating this reaction. $Zn^{2+}$ has the opposite effect, decreasing the pH by lowering the pK$_a$ of Zn$^{2+}$-bound water molecules, and thereby inhibiting the Mn$^{2+}$ oxidation reaction that would otherwise occur. When base is added to this mixture during the synthesis of Mn$^{2+}$:ZnO nanocrystals, we also do not observe the appearance of an absorption shoulder at 20000 cm$^{-1}$ (Figure 1), and it is reasonable to assume that the small amount of Mn$^{2+}$ present (in comparison with Zn$^{2+}$) is readily incorporated into basic zinc acetate clusters as they form, and this prohibits its coalescence into manganese-rich phases. Incorporation of Co$^{2+}$ into basic zinc acetate clusters was also concluded from a combination of absorption spectra and nucleation inhibition data.[32] The conclusion that manganese oxides are not formed is supported by the observation that all of the manganese present in the final product can be accounted for as paramagnetic $S = 5/2$ Mn$^{2+}$ (Section IV.H), whereas manganese-rich phases would show exchange-dominated magnetism. In summary, the formation of phase-segregated manganese oxides appears to be prevented very effectively by the high concentration of Zn(OAc)$_2$ also present in the reaction mixture. This inhibition allows high-quality Mn$^{2+}$:ZnO to be synthesized under aerobic conditions without addition of reductants to combat manganese oxidation. These conclusions are summarized in Scheme 1.

**Scheme 1.**

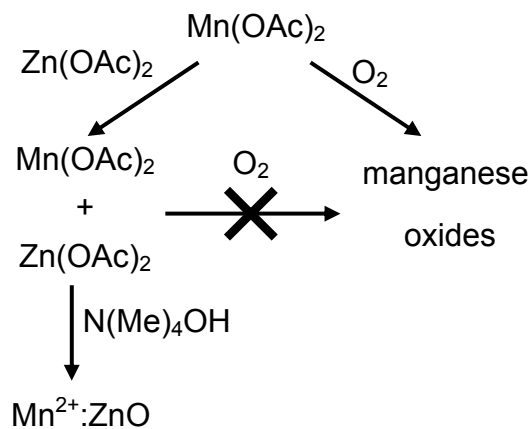

**F. Charge-transfer transitions in Mn$^{2+}$:ZnO.** The absorption spectra of Mn$^{2+}$:ZnO colloids and thin films show only one feature in addition to the absorption of ZnO, namely a broad tailing sub-bandgap absorbance at ca. 24000 cm$^{-1}$ (Figures 1 and 6). Although similar broad absorption occurs in many manganese oxides, this feature is unambiguously associated with magnetically dilute Mn$^{2+}$ ions. This conclusion is drawn on the basis of the MCD spectroscopy of the freestanding $Mn^{2+}$:ZnO nanocrystals shown in Figure 6, in which the broad absorption feature is seen to give rise to a similarly broad negative pseudo-𝒜 term MCD signal centered at 23600 cm$^{-1}$. The 5 K MCD intensity shows $S = 5/2$ saturation magnetization (Figure 6), demonstrating definitively that this intensity arises from an $S = 5/2$ chromophore and hence from magnetically isolated $Mn^{2+}$ ions. The negative MCD signal to higher energy of the pseudo-𝒜 signal is the onset of the first excitonic transition, which gains MCD intensity primarily through the covalency of the $Mn^{2+}$-oxo bonds (p-d exchange).[1] The first excitonic transition is subject to so-called "giant Zeeman splittings",[1] and the large shift in its energy with applied magnetic field results in nonsuperimposable MCD spectra at different fields in this energy region. Quantitative analysis of this Zeeman shift is currently being investigated and will be reported separately.

Broad sub-bandgap absorption features similar to the one we observe have been reported previously for other $Mn^{2+}$:ZnO preparations,[17,22,37,38] and have been assigned variously as $Mn^{2+}$ ligand field[22,37,38] or charge-transfer (CT)[17] transitions. Using the high-quality colloids prepared here, the molar extinction coefficient (per $Mn^{2+}$) of this feature was determined accurately to be $\varepsilon_{Mn^{2+}} = 950$ M$^{-1}$cm$^{-1}$ at 24000 cm$^{-1}$ and 300 K. This sub-bandgap absorption feature is approximately two orders of magnitude more intense than could be reasonably expected for $Mn^{2+}$ ligand-field transitions ($\varepsilon_{Mn^{2+}}(^6A_1 \rightarrow {}^4T_1(G)) = 1$-$10$ M$^{-1}$cm$^{-1}$ in tetrahedral coordination complexes[47] and in ZnS[48]). Furthermore, the 5 K MCD spectrum in this region is structureless (Figure 6), in contrast with what would be observed were this intensity to arise from the closely spaced $^6A_1 \rightarrow {}^4T_1(G), {}^4T_2(G), {}^4A_1(G), {}^4E(G)$ series of $Mn^{2+}$ ligand-field transitions expected to occur in this same energy region (see Supporting Information). Structured diffuse reflectance spectra in this energy region have been reported[22] but only at relatively high $Mn^{2+}$ concentrations where dimer exchange effects may contribute to relaxation of the spin selection rules, and similar structure was not observed in thin films prepared at lower $Mn^{2+}$ concentrations.[17,37,38] The intensity and bandshape of this sub-bandgap absorption feature are thus inconsistent with its identification as $Mn^{2+}$ ligand field transitions, and suggest it should be assigned as CT intensity. We consider two classes of CT transitions that may occur in $Mn^{2+}$:ZnO. In one case, an electron may be promoted from the $Mn^{2+}$ ion to ZnO-based acceptor orbitals of the conduction band (CB), and in the other case an electron may be promoted to the $Mn^{2+}$ ion from ZnO-based donor orbitals of the valence band (VB). The state generated at the CT electronic origin is expected to involve a semiconductor electron (or hole) that is loosely bound to the charged impurity by Coulombic forces,[49] although possibly with a large effective radius. Treating the semiconductor nanocrystal as a ligand to the dopant ion, these transitions are formally analogous to metal-to-ligand CT (MLCT) and ligand-to-metal CT (LMCT) transitions. Increasing the excitation energy may be sufficient to induce photoionization, in which the electron or hole is in an unbound conduction or valence band level. Although binding energies for $Mn^{2+}$ ions in II-VI semiconductors are not known, typical binding energies for the VB → $TM^{2+}$ CT excited states in $Ni^{2+}$-doped II-VI semiconductors are ca. 250 $cm^{-1}$.[49]

The intensity of the CT transition observed in the $Mn^{2+}$:ZnO colloids of Figure 1b ($\varepsilon_{Mn^{2+}} \approx 950$ $M^{-1}cm^{-1}$ at 24000 $cm^{-1}$) is comparable to that of the acceptor transition in $Ni^{2+}$:ZnO ($\varepsilon_{Ni^{2+}} \approx 700$ $M^{-1}cm^{-1}$ at 23100 $cm^{-1}$),[32,50] suggesting that it may also be a VB → $TM^{2+}$ CT transition. Because CT absorption intensities scale with donor-acceptor covalency,[51] and s-d (CB-$Mn^{2+}$) hybridization is generally an order of magnitude smaller than p-d (VB-$Mn^{2+}$) hybridization in II-VI DMSs,[1,7] $TM^{2+}$ → CB CT intensities are expected to be considerably weaker than VB → $TM^{2+}$ CT transitions in the absorption spectra of $Mn^{2+}$:ZnO and other II-VI DMSs. The observed transition occurs much lower in energy than anticipated for a LMCT transition involving $Mn^{2+}$, however, casting some uncertainty on this assignment.

Charge-transfer transition energies can be analyzed in the context of Jørgensen's optical electronegativity model,[52] applied previously for analysis of similar broad features observed in the MCD and absorption spectra of $Co^{2+}$:ZnO and $Ni^{2+}$:ZnO DMSs.[32] Following Jørgensen, Equation 3 relates the energy of a CT transition, $E_{CT}$, to the donor (D) and acceptor (A) electronegativities ($\chi$), taking into account differences in spin-pairing energies (SPE) between the ground and excited electronic states.[47,52]

$$E_{CT} = 30000 cm^{-1} \cdot \left[ \chi_{opt}(D) - \chi_{opt}(A) \right] + \Delta SPE \pm 10Dq \qquad (3)$$

ΔSPE for $Mn^{2+}$ acceptor ($d^5 \rightarrow d^6$) and donor ($d^5 \rightarrow d^4$) CT transitions are both $+28/9[(5/2)B + C]$,[47] where $B$ is the Racah electron-electron repulsion parameter. Reliable ligand field parameters are not available for $Mn^{2+}$ in ZnO, and in general are difficult to ascertain for $Mn^{2+}$ since the energies of the observable ligand field excited states depend largely on electron-electron repulsion magnitudes. For our analysis, we have estimated $Dq = 420$ cm$^{-1}$, $B = 596$ cm$^{-1}$, and $C/B = 6.5$ by extrapolation of ligand field parameters determined previously for $Mn^{2+}$, $Co^{2+}$, and $Ni^{2+}$ tetrahalide complexes[47] and $Co^{2+}$- and $Ni^{2+}$-doped ZnO[50] (see Supporting Information). Although $Mn^{2+}$ can accept or donate electrons using either the $e$ or $t_2$ d-orbitals, correction of Equation 3 for the tetrahedral ligand-field splitting (10Dq) is likely necessary because in both cases the CT absorption intensity is dominated by $Mn^{2+}$-ZnO covalency predominantly involving the $t_2$ d orbitals. From absorption spectra of octahedral $MnX_2$ (X = Cl, Br), $\chi_{opt}(Mn^{2+}(O_h))$ has been estimated to be 1.45.[53] The value of $\chi_{opt}(Mn^{2+}(T_d))$ is expected to be slightly larger,[47] but for the current analysis it is sufficient to use the approximate value of 1.45. The optical electronegativities of the valence and conduction bands in pure ZnO have been estimated previously to be $\chi_{opt}(CB) \approx 1.1$ and $\chi_{opt}(VB) \approx 2.0$.[53] Using these values, solution of Equation 3 yields predicted energies of $E_{LMCT} \approx 37400$ cm$^{-1}$ and $E_{MLCT} \approx 23000$ cm$^{-1}$. Using the relatively well defined electronegativities of tetrahedral $Co^{2+}$ and $Ni^{2+}$ ions, $\chi_{opt}(VB) \approx 2.4$ was determined from $Ni^{2+}$:ZnO and $Co^{2+}$:ZnO VB → $TM^{2+}$ CT transition energies,[32] and with this value we predict $E_{LMCT} \approx 49400$ cm$^{-1}$. The high energies predicted for the VB → $Mn^{2+}$ CT transition in $Mn^{2+}$:ZnO are consistent with the absence of LMCT transitions in $MnX_4^{2-}$ (X = Cl, Br) below 38000 cm$^{-1}$.[47] Because of the high LMCT energies of these analogous tetrahedral complexes and predicted for $Mn^{2+}$ in ZnO, VB → $Mn^{2+}$ CT absorption intensity at 24000 cm$^{-1}$ in $Mn^{2+}$:ZnO appears improbable. Alternatively, the predicted $TM^{2+}$ → CB CT energy ($E_{MLCT} \approx 23000$ cm$^{-1}$) is in reasonable agreement with the experimental energy (ca. 24000 cm$^{-1}$, Figures 1 and 6), and we tentatively assign this absorption intensity as due to a $Mn^{2+} \rightarrow$ CB excitation. Interestingly, photocurrent action spectra of $Mn^{2+}$:ZnO thin films reveal photocurrents generated with photon energies down to 14500 cm$^{-1}$.[38] This activity was attributed to $Mn^{2+}$ ligand field absorption, but appears to arise from the same transitions observed in the spectra of Figure 6. The photocurrent action begins ca. 10000 cm$^{-1}$ lower in energy than the first Mn$^{2+}$ ligand field excited state ($^4T_1$(G)), predicted at ca. 24900 cm$^{-1}$ from the Tanabe-Sugano matrices using the ligand field parameters listed above (see Supporting Information). The lower energy of the Mn$^{2+}$ → CB CT excitation than the $^4T_1$(G) Mn$^{2+}$ ligand-field excitation has important consequences for the luminescence properties of Mn$^{2+}$:ZnO, as described in the following section (Section IV.G).

**G. Luminescence of Mn$^{2+}$:ZnO nanocrystals.** ZnO nanocrystals are known for their characteristic green luminescence, which originates from e$^-$-h$^+$ recombination involving surface trap states.[54] Figure 7 shows the 300K luminescence spectrum of undoped colloidal ZnO nanocrystals prepared by the same procedure that was used to make the Mn$^{2+}$-doped nanocrystals reported here. The green emission, which peaks at about 18600 cm$^{-1}$, is the same as that reported previously.[54] In addition, a pronounced UV excitonic emission peak is observed from these colloids. The large ratio of UV:green emission in Figure 7 is not observed in TOPO-capped ZnO nanocrystals prepared by an analogous procedure[32] or by hot injection,[55] and it therefore arises specifically from the use of dodecylamine as the capping ligand, presumably by surface passivation. A similar enhancement of excitonic emission was reported for ZnO colloids capped with polyvinyl pyrrolidone ligands.[56]

Mn$^{2+}$-doped II-VI semiconductor nanocrystals also commonly show emission from the Mn$^{2+}$ $^4T_1$(G) ligand-field excited state. This emission has been observed in Mn$^{2+}$:ZnS,[25] Mn$^{2+}$:ZnSe,[27] and Mn$^{2+}$:CdS[26] quantum dots at about 17200 cm$^{-1}$ (580 nm) with a considerably smaller bandwidth than the surface trap emission. The two Mn$^{2+}$:ZnO nanocrystal samples in Figure 7 show no evidence of Mn$^{2+}$ emission, however, but show only quenching of the ZnO excitonic and surface trap emission. The surface trap emission of the pure ZnO QDs is reduced by ca. 42% with 0.13% Mn$^{2+}$ doping, while the excitonic emission is quenched to a slightly greater extent. At this Mn$^{2+}$ concentration, each ca. 6.5 nm diameter nanocrystal contains an average of approximately 8 Mn$^{2+}$ ions (see Supporting Information for statistics). The observation of any ZnO surface trap emission at all in these doped nanocrystals therefore indicates that Mn$^{2+}$ is not a particularly effective trap. Almost complete quenching of the surface trap emission is observed with the modest dopant concentration of 1.3% (Figure 7), however. In no instance was Mn$^{2+}$ ligand field emission observed. Quenching of the ZnO luminescence by Mn$^{2+}$ and the lack of Mn$^{2+}$ ligand-field emission in ZnO have both been noted in previous studies of bulk Mn$^{2+}$:ZnO.[57] These observations are explained by the conclusion drawn in Section IV.F that the threshold of the MLCT transition lies below the Mn$^{2+}$ $^4T_1(G)$ state in ZnO. This low-lying MLCT level would provide a pathway for non-radiative decay of the excited Mn$^{2+}$ ions. The relatively localized wavefunctions of the surface trap and Mn$^{2+}$ electronic states imply that direct quenching of surface trap emission by Mn$^{2+}$ should proceed by Förster energy transfer, but the extremely low oscillator strengths of the Mn$^{2+}$ visible absorption transitions indicate that Mn$^{2+}$ dopants cannot quench surface traps effectively by this mechanism. Rather, it is likely that Mn$^{2+}$ reduces surface trap emission by competing with surface states in trapping the ZnO excitonic excitation energy, a conclusion supported by the observation that excitonic emission is quenched somewhat more effectively than surface trap emission in the 0.13% Mn$^{2+}$:ZnO nanocrystals shown in Figure 7.

**H. High-$T_C$ ferromagnetism in Mn$^{2+}$:ZnO thin films.** The EPR (Figure 5) and magnetic susceptibility (Figures 8 and 9b) data collected for the freestanding nanocrystals show only the paramagnetic phase of the DMS Mn$^{2+}$:ZnO. Neglecting the small zero-field splitting of the Mn$^{2+}$ $^6A_1$ ground state ($|D|$ = 2.36x10$^{-2}$ cm$^{-1}$), the 5 K magnetization of freestanding 0.20 ± 0.01% Mn$^{2+}$:ZnO nanocrystals can be reproduced quantitatively using the Brillouin function (Equation 4)[58] and the experimental parameters determined from the EPR simulations discussed in Section IV.C ($S$ = 5/2 and $g$ = 1.999), with no fitting.

$$M = \frac{1}{2} Ng\mu_B \left[ (2S+1)\coth\left((2S+1)\left(\frac{g\mu_B H}{2kT}\right)\right) - \coth\left(\frac{g\mu_B H}{2kT}\right) \right] \quad (4)$$

Here, $\mu_B$ is the Bohr magneton and N represents the number of Mn$^{2+}$ ions in the sample, derived from the experimental Mn$^{2+}$ concentrations determined by ICP-AES. The predicted 5 K magnetization is superimposed on the experimental data for the freestanding nanocrystals in Figure 9b and the two are identical within experimental error bars (arising from the error bars in Mn$^{2+}$ concentration). The quantitative agreement between the experimental and calculated magnetization demonstrates that all of the manganese in these nanocrystals is accounted for as paramagnetic $Mn^{2+}$ and there is little or no influence from antiferromagnetic superexchange interactions at this low dopant concentration. As mentioned in Section IV.E, this result demonstrates that no significant phase segregation of manganese oxides has taken place during synthesis.

When these nanocrystals are spin coated into thin films their magnetic properties change dramatically. Films A, B, and C all exhibit robust ferromagnetism at room temperature (Figure 8), with relatively minor changes in hysteresis properties over the entire accessible temperature range ($\leq$ 350 K, Figure S2). The 300 K saturation moment of $M_S$ = 1.35 $\mu_B/Mn^{2+}$ for Film C exceeds the only other reported 300 K $Mn^{2+}$:ZnO ferromagnetic saturation moment ($M_S$ = 0.16 $\mu_B/Mn^{2+}$)[15] by nearly an order of magnitude, and is similar to the 10 K $M_S$ value reported recently for a manganese-doped GaAs thin film grown by molecular beam epitaxy and applied in a spin-LED device.[9]

The ferromagnetism probed by magnetic susceptibility (Figures 8 and 9) also manifests itself in the EPR spectroscopy of these thin films (Figure 10). The 300K EPR spectrum of the freestanding 0.20% $Mn^{2+}$:ZnO colloids used to make Film A (Figure 10, dashed) shows the sharp features centered at $g$ = 1.999 analyzed in Figure 5 for nominally 0.02% $Mn^{2+}$:ZnO nanocrystals. In contrast, the 300K EPR spectrum of Film A exhibits a broad resonance spanning the entire field range in Figure 10. This broad feature, a so-called ferromagnetic resonance (FMR) signal,[59] arises from transitions within the ground state of a ferromagnetic domain. Its breadth likely arises in part from the intrinsic anisotropy of the signal. For example, separations of as large as 5000 Oe are observed between parallel and perpendicular resonance fields in oriented manganese-doped GaAs thin films,[59] and such large anisotropies could result in broad spectra for powder samples. The breadth may also arise in part from the very high multiplicity of the ferromagnetic domain's ground state due to its high effective spin state (estimated from the 300 K saturation magnetization curvature in Figure 8 to exceed $S$ = 800 on average). Finally, inhomogeneity in domain sizes may contribute to the breadth of the signal. A similar 300 K FMR signal was reported recently for nominally 2% $Mn^{2+}$:ZnO bulk powders prepared by high-temperature solid-state fusion[15] and is also observed in ferromagnetic $Co^{2+}$:ZnO thin films prepared by MOCVD.[60]

Using the high-quality colloidal $Mn^{2+}$:ZnO nanocrystals as solution precursors for spin coating ensures an even distribution of $Mn^{2+}$ ions throughout the resulting $Mn^{2+}$:ZnO thin film. XRD data collected for Film A (Figure 3a) show narrower diffraction peaks than were observed for the paramagnetic nanocrystals. Analysis of the XRD peak widths using the Scherrer equation indicates an increase in effective crystal diameters from 6 to 20 nm, consistent with nanocrystal sintering. There is no evidence of any phase segregation in the XRD data, which show excellent signal-to-noise ratios, although this method would likely be too insensitive to detect phase segregation at such low manganese concentrations. The solid solubility of $Mn^{2+}$ in ZnO is high, however, exceeding 10% at 525 °C and 1 kbar,[22] and $Mn^{2+}$ phase segregation during the brief annealing of these 0.20% $Mn^{2+}$:ZnO nanocrystals at 525 °C is therefore considered extremely unlikely. ZFC magnetization data have been measured for Films A and B and the two data sets are nearly identical (Figure 9a). Both show increasing magnetization as the temperature is elevated, indicative of spontaneous cooperative magnetization. Notably, no magnetic phase transitions are observed in the ZFC data, confirming the absence of MnO or $Mn_3O_4$ phase segregated impurities, nanocrystals of which exhibit magnetic phase transitions below ca. 45 K that show up as pronounced maxima in ZFC magnetization measurements.[61]

In addition to nearly temperature-independent ferromagnetism, the films show residual paramagnetism that follows Curie behavior in both field and temperature. The paramagnetic contribution in Film A was quantified by subtracting the ferromagnetic signal from the 5 K data in Figure 8 (plotted in Figure 9b), and from the 1 Tesla data at various temperatures (plotted in Figure 9c). Both data sets were then fitted using the Brillouin function (Equation 4) with only one floating parameter, N, the number of paramagnetic $Mn^{2+}$ ions. The two data sets agree quantitatively and are best fit when $N = 0.64\ N_{tot}$, where $N_{tot}$ is the total manganese content of the film (0.20 ±0.01%). The paramagnetic moment per $Mn^{2+}$ in Film A has thus been reduced to ca. 64% of its value in the freestanding nanocrystals (Figure 9b), and the other 36% of the $Mn^{2+}$ ions have undergone a magnetic phase transition. A lower limit for the number of $Mn^{2+}$ ions participating in the ferromagnetic domains can be estimated from the ferromagnetic saturation moments by assuming a maximum moment of 5 $\mu_B/Mn^{2+}$. With this approximation, 13%, 24%, and 27% of the $Mn^{2+}$ ions are ferromagnetically aligned at 300 K for Films A, B, and C, respectively. These numbers represent lower limits, because the apparent moments per $Mn^{2+}$ will be reduced by the contributions of carriers, antiferromagnetic exchange interactions, or spin-glass behavior. We speculate, for example, that at least some $Mn^{2+}$ ions are antiferromagnetically exchange coupled to the ferromagnetic domains as observed previously in $Ni^{2+}$:ZnO.[24]

The EPR signal of paramagnetic $Mn^{2+}$ in ZnO can be observed superimposed on the broad FMR signal of Film A in Figure 10, consistent with the presence of substantial paramagnetic $Mn^{2+}$ in these films as concluded from analysis of the magnetic susceptibility data (Figure 9). This paramagnetic EPR signal was not observed in the EPR spectrum of ferromagnetic $Mn^{2+}$:ZnO reported previously,[15] despite the fact that less than 4% of the manganese in those samples could be accounted for as ferromagnetic ($M_S$ = 0.16 $\mu_B/Mn^{2+}$), leaving more than 96% of the manganese unaccounted for. The EPR signal of paramagnetic $Mn^{2+}$ is exceptionally intense at room temperature, and its absence combined with the weaker ferromagnetism in the material reported previously may suggest that a substantial portion of the manganese in that material remained antiferromagnetically coupled within $MnO_2$ crystallites, since $MnO_2$ was the original source of manganese used in the synthesis. The stronger ferromagnetism observed in the films prepared from the colloidal doped nanocrystals, coupled with the observation of residual paramagnetic $Mn^{2+}$, are both likely the consequence of a more homogeneous distribution of substitutional $Mn^{2+}$ dopants throughout the material.

Finally, a new, sharp resonance at $g$ = 2.00 (3362 Gauss) is observed in the EPR spectrum of the thin film (Figure 10) that was not present in the freestanding colloids. This sharp feature resembles a radical EPR signal and suggests that redox chemistry is occurring during preparation of the thin films. Ferromagnetism in ZnO DMSs is widely believed to be carrier mediated.[2,3] Zener model calculations[2] and local-density-approximation density functional theory (LDA-DFT) calculations[3] both predict that ferromagnetism in $Mn^{2+}$:ZnO requires the presence of relatively large carrier concentrations ($p \approx 3.5 \times 10^{20}$ cm$^{-3}$). A sensitive dependence on carrier concentration would be consistent with the observation that $Mn^{2+}$:ZnO prepared by similar methods can show substantially different magnetic properties[10,15-18] (see also Supporting Information for data on additional films from this study). Importantly, the activating carriers in both models are p-type. In contrast with III-V DMSs such as $Mn^{2+}$:GaAs, substitutional doping of $TM^{2+}$ ions into ZnO does not itself generate p-type carriers, so these carriers must be introduced by other routes. The radical-like feature we observe in the EPR spectrum of the ferromagnetic $Mn^{2+}$:ZnO thin film suggests that redox chemistry during film preparation could possibly be the source of the required carriers. ZnO forms n-type defects under growth conditions very readily, however, and so it would be surprising if radical chemistry could introduce sufficient p-type defects in our films to explain our magnetic data within the existing theoretical models. Indeed, growth of p-type ZnO has been a long-standing challenge because of the ubiquity of compensating n-type defects.[62] Only relatively recently was p-type ZnO successfully prepared,[63] by introducing NO, $NH_3$, $N_2$, or $N_2O$ gasses during ZnO growth to incorporate N heteroatoms at anion sites of the wurtzite lattice. One intriguing possibility is that calcination of the spin-coated dodecylamine-ligated $Mn^{2+}$:ZnO nanocrystals leaves behind nitrogen heteroatoms that serve as p-type defects by a process analogous to those reported previously.[63] Further investigation is needed to fully unravel the issues of magnetic ordering mechanism and carrier type in ferromagnetic $Mn^{2+}$:ZnO, but the possibility that the chemical identity of the capping ligand may play an important role when starting from nanocrystalline precursors exposes the exciting possibility that room-temperature magnetic ordering in $Mn^{2+}$:ZnO and related DMSs may ultimately be controlled, and hence understood, using chemical perturbations. Experiments exploring these subjects are currently underway.

### V. Conclusion

We have demonstrated the preparation of high-quality colloidal $Mn^{2+}$-doped ZnO diluted magnetic semiconductor quantum dots by a direct solution chemical route. Segregation of dopant phases is a major concern in the synthesis of DMSs by any method because it may interfere with observation of the intrinsic properties of the target DMS. In many ZnO DMS preparations reported previously, methods have been used that may actually promote dopant segregation through high temperatures or reductive conditions, or through difficulties in controlling the form in which dopants are introduced. We have demonstrated that it is possible to prepare colloidal $Mn^{2+}$:ZnO nanocrystals with extremely homogeneous dopant speciation, in which all of the $Mn^{2+}$ is substitutionally doped within the cores of the nanocrystals and not on the crystal surfaces, by direct chemical synthesis from homogeneous solution in air at room temperature. Preparation of thin films by spin-coat processing using these colloids as solution precursors yields strongly ferromagnetic $Mn^{2+}$:ZnO thin films with Curie temperatures well above room temperature and with 300K saturation moments up to 1.35 $\mu_B/Mn^{2+}$, nearly one order of magnitude greater than the only previously reported room-temperature value for $Mn^{2+}$:ZnO.[15] These results demonstrate the successful application of direct chemical routes to the preparation of a strongly ferromagnetic semiconductor that has been predicted to play an important role in the emerging field of spin-based electronics technologies. The insights gained from chemical experiments with this and other diluted magnetic semiconductors are expected to help guide the preparation of increasingly high-quality ferromagnetic semiconductors and improve the fundamental understanding of their interesting physical properties.

**Acknowledgment.** This work was funded by the NSF (DMR-0239325 and ECS-0224138). The authors are grateful to the UW/PNNL Joint Institutes for Nanoscience for graduate support (N.S.N., K.R.K.), to Prof. Høgni Weihe (University of Copenhagen) and Prof. Philip Tregenna-Piggott (University of Bern) for invaluable assistance with the EPR simulations, and to Dr. Chongmin Wang (PNNL) and Dr. J. Daniel Bryan (UW) for assistance with TEM measurements. NIH Center grant P30 ES07033 is acknowledged for supporting the X-band EPR instrument at UW. Q-band EPR and TEM measurements were performed at EMSL, a national user facility sponsored by the U.S. DOE's Office of Biological and Environmental Research located at PNNL and operated by Battelle. D.R.G. is a Cottrell Scholar of the Research Corporation.

**Supporting Information Available:** Three tables: A compilation of literature data used to estimate ligand-field parameters and excited-state energies for $Mn^{2+}$ in ZnO. Three figures: (**S1**) Doping statistics for $Mn^{2+}$:ZnO nanocrystals. (**S2**) Temperature dependence of magnetic hysteresis parameters for Films A - C. (**S3**) 300 K magnetic data for additional $Mn^{2+}$:ZnO nanocrystalline thin films. This material is available free of charge via the Internet at http://pubs.acs.org.

**References.**

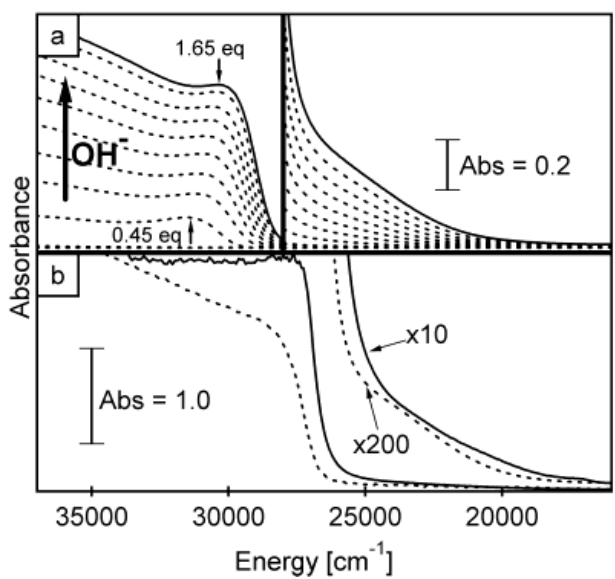

**Figure 1.** (a) 300 K absorption spectra of a 0.10 M DMSO solution of 2% Mn(OAc)$_2$/98% Zn(OAc)$_2$ collected following successive additions of 0.15 equiv of 0.55 M N(Me)$_4$OH in ethanol. The solid line shows the data collected after 1.65 equiv of base were added. (b) 300 K absorption spectra of (----) colloidal 0.20% Mn$^{2+}$:ZnO nanocrystals after treatment with dodecylamine and (—) a thin film prepared by spin coating the same nanocrystals onto fused silica (Film A). The thin film absorbance reaches the stray light limit at ca. 27000 cm$^{-1}$.

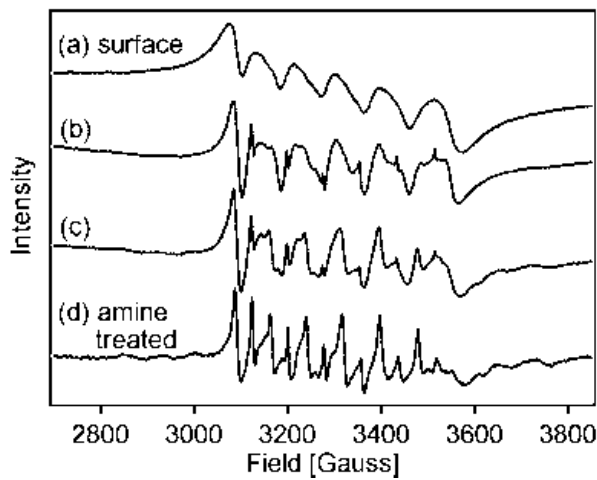

**Figure 2.** 300K X-band EPR spectra of colloidal Mn$^{2+}$:ZnO nanocrystals. (a) Surface-bound Mn$^{2+}$:ZnO nanocrystals. Samples prepared from 0.02% Mn$^{2+}$/99.98% Zn$^{2+}$ reaction solution collected (b) 10 minutes after base addition, (c) after 2 hours of heating at 60 °C, and (d) after treating with dodecylamine.

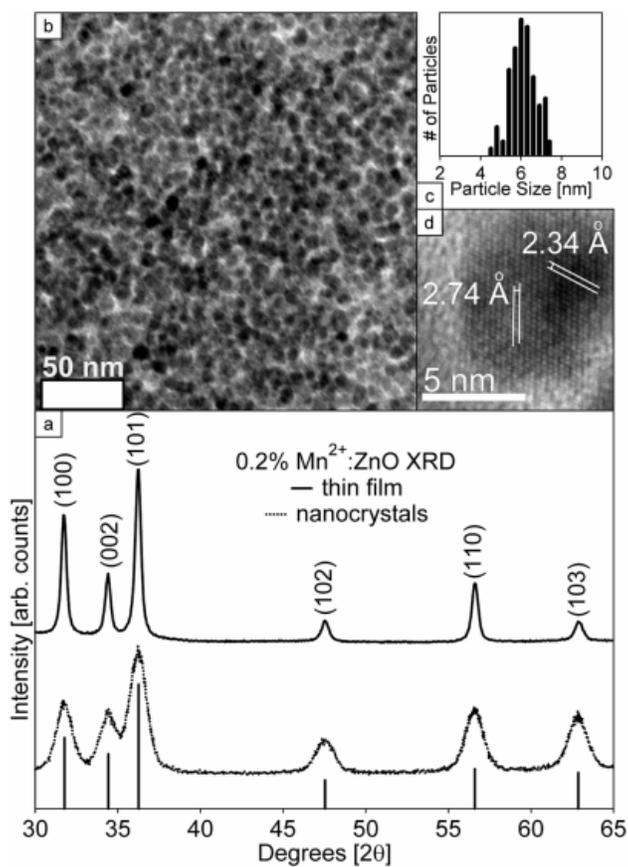

**Figure 3.** (a) Powder X-ray diffraction data: (···) 0.20% $Mn^{2+}$:ZnO nanocrystals. (—) thin film prepared from the 0.20% $Mn^{2+}$:ZnO nanocrystals. Peak positions for wurtzite ZnO are included for reference. (b) Overview TEM image of 0.20% $Mn^{2+}$:ZnO nanocrystals. (c) Histogram of 100 crystal diameters (6.1 ± 0.7 nm average diameter). (d) High-resolution TEM image of a single nanocrystal showing lattice spacings matching those of ZnO.

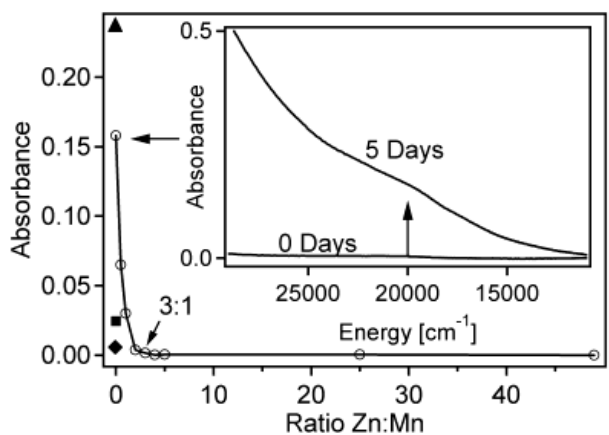

**Figure 4.** Absorption intensities of various solutions of 0.002 M Mn(OAc)$_2$ in DMSO measured at 20000 cm$^{-1}$ after 5 days in air. (○) Solutions with varying ratios of Zn(OAc)$_2$ to Mn(OAc)$_2$. (▲) A solution with a 5:1 ratio of Na(OAc) to Mn(OAc)$_2$. (■) An anaerobic solution of Mn(OAc)$_2$. (♦) A solution with 0.002 M Mn(NO$_3$)$_2$ instead of Mn(OAc)$_2$. Inset: Absorption spectra of a DMSO solution of 0.002 M Mn(OAc)$_2$ after mixing (0 days) and after 5 days in air.

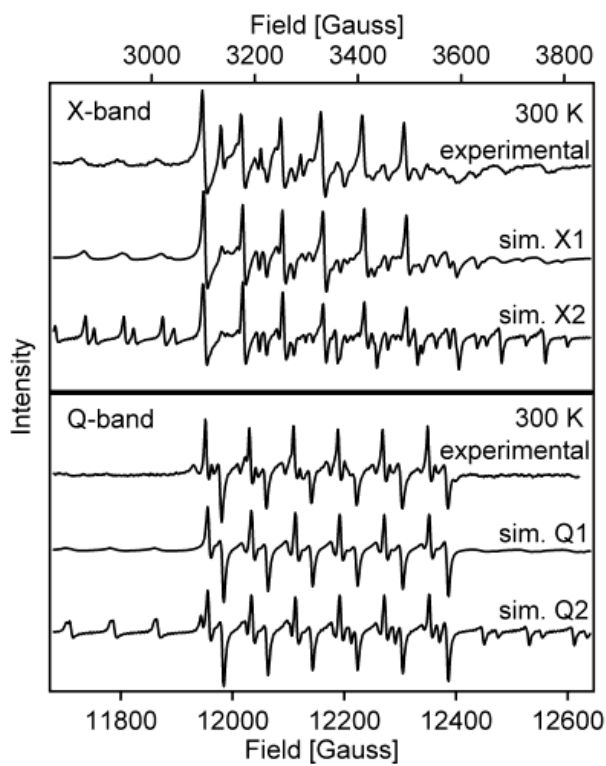

**Figure 5.** Experimental and simulated 300 K X- and Q-band EPR spectra of colloidal dodecylamine-capped 0.02% Mn$^{2+}$:ZnO nanocrystals in toluene. Simulations with (X1 and Q1) and without (X2 and Q2) σ = 2% $D$-strain are included.

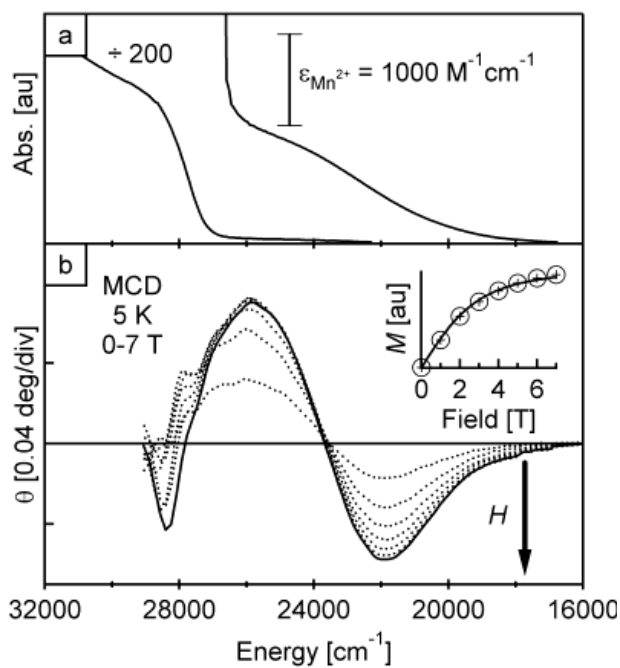

**Figure 6.** (a) 300 K absorption spectrum of colloidal TOPO-capped 1.1% $Mn^{2+}$:ZnO nanocrystals. (b) Variable-field 5 K MCD spectra of the same nanocrystals, drop-coated onto quartz discs to form frozen solutions. The inset shows the 5 K MCD saturation magnetization probed at 21900 $cm^{-1}$. The solid line shows the $S = 5/2$ saturation magnetization at 5 K predicted by the Brillouin function (Equation 4).

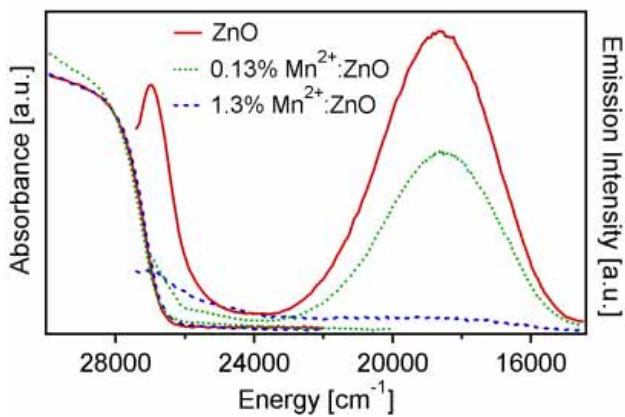

**Figure 7.** 300 K absorption and luminescence spectra of pure ZnO (—), 0.13% $Mn^{2+}$:ZnO (estimated concentration) (···), and 1.3% $Mn^{2+}$:ZnO (---) colloids.

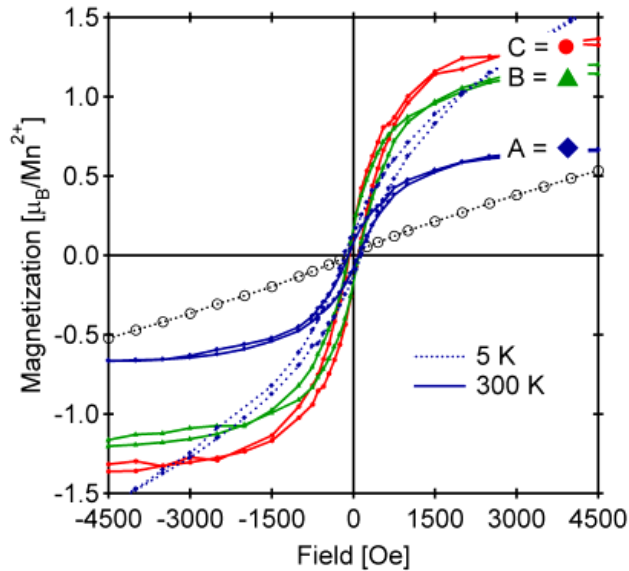

**Figure 8.** 5 K (⋯) and 300 K (—) magnetic susceptibilities of 0.20% $Mn^{2+}$:ZnO nanocrystals (○) and thin films (A = ◆, B = ▲, C = ●).

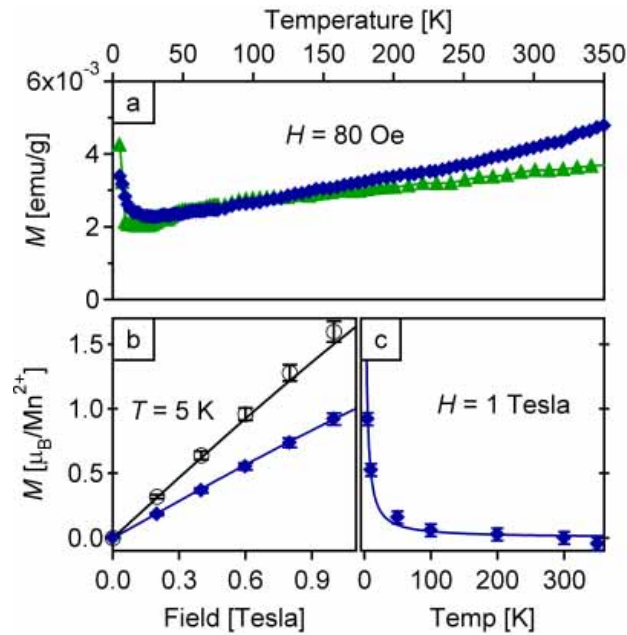

**Figure 9.** Magnetic susceptibility data for 0.20(±0.01)% $Mn^{2+}$:ZnO. (a) Zero-field-cooled magnetization of Films A (◆) and B (▲) with an applied magnetic field of 80 Oe. (b) 5 K field-dependent magnetization of freestanding nanocrystals (○) and of the residual paramagnetic magnetization of Film A. (c) Temperature dependence of the residual paramagnetic magnetization of Film A with $H = 1$ T. The solid lines in (b) and (c) show the Brillouin function calculated using $S = 5/2$ and $g = 1.999$ (Equation 4, see text). The error bars represent the uncertainty in $Mn^{2+}$ concentration.

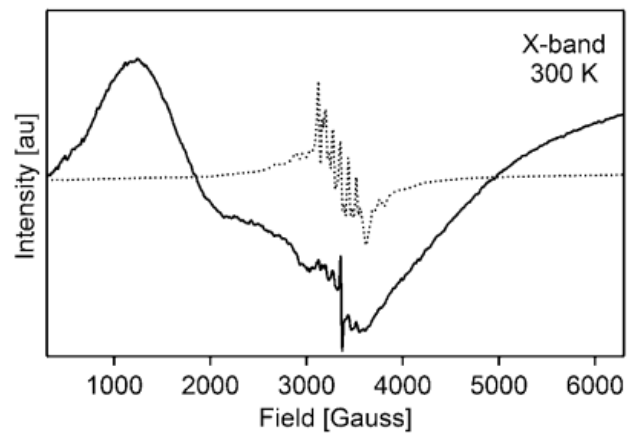

**Figure 10.** 300 K X-band EPR spectra: (—) Ferromagnetic 0.20% Mn$^{2+}$:ZnO thin film (Film A) from Figure 8. (···) The colloids from which Film A was prepared, suspended in toluene.